\documentclass{article}
\usepackage[utf8]{inputenc}
\usepackage{graphicx}
\usepackage{appendix}
\usepackage{enumitem}
\usepackage{amsmath}
\graphicspath{ {images/} }

\author{Max Henderson, Jarred Gallina, Michael Brett}
\date{April 2020}

\begin{document}

\title{Methods for Accelerating Geospatial Data Processing Using Quantum Computers}
\maketitle

\begin{abstract}
Quantum computing is a transformative technology with the potential to enhance operations in the space industry through the acceleration of optimization and machine learning processes. Machine learning processes enable automated image classification in geospatial data. New quantum algorithms provide novel approaches for solving these problems and a potential future advantage over current, classical techniques. Universal Quantum Computers, currently under development by Rigetti Computing and other providers, enable fully general quantum algorithms to be executed, with theoretically proven speed-up over classical algorithms in certain cases. This paper describes an approach to satellite image classification using a universal quantum enhancement to convolutional neural networks: the quanvolutional neural network. Using a refined method, we found a performance improvement over previous quantum efforts in this domain and identified potential refinements that could lead to an eventual quantum advantage.

We benchmark these networks using the SAT-4 satellite imagery data set in order to demonstrate the utility of machine learning techniques in the space industry and the potential advantages that quantum machine learning can offer.

\end{abstract}

\section{Introduction}

\subsection{Quantum Computing for Aerospace Applications}

From solving systems of nonlinear equations to
processing massive amounts of “big data”, the aerospace field is rife with computational challenges. We explore here the problem of image classification using lower-resolution data. Many applications within aerospace depend on reliable satellite imagery, which has historically been an expensive resource to collect and utilize. However, the entry of small satellite operators to the market has increased the supply of lower cost imagery. Although this data is of lower resolution than that from more traditional providers, powerful post-processing techniques bolster the utility of the data, making its usefulness rival their more expensive counterparts for certain applications. For example, techniques for deriving sub-pixel details from lower resolution imagery have been applied in a
variety of use cases like crop mapping \cite{Atzberger2013MappingTS}. Post-processing techniques that leverage machine learning
algorithms garner information from low-resolution
data that previous techniques cannot. However, these algorithms can still be challenging, costly and time consuming. One successful but \mbox{computationally-expensive} technique is convolutional neural
networks (CNNs) \cite{cnn}. CNNs are a class of neural
networks commonly used to analyze visual imagery. In
order to address overfitting, they are more regularised
than deep belief networks, and prioritise correctness
over complexity.
One avenue for improvement of CNNs is to look to
the field of Quantum Machine Learning (QML) \cite{biamonte2017quantum}.

\subsection{Introduction to Quantum Computing}
Quantum computing is a novel technology that has
seen an explosion in academic and industrial interest.
Quantum computers are devices composed of quantum
systems (typically quantum bits, or qubits) that leverage
fundamental quantum mechanical phenomena, such as
superposition, entanglement, and tunnelling. These
quantum phenomena allow for the development of
completely new types of “quantum” algorithms, some of
which have been shown to be exponentially more
efficient than any known classical computing
counterpart \cite{Shor_1997}. Such promising results have spurred
the development of improved quantum hardware
alongside investigations into quantum computing
algorithms for application spaces with highly important
and \mbox{classically-intractable} problems. Solving aerospace big data problems using quantum computation has
become a matter of focus for NASA \cite{Rieffel2019, Smelyanskiy2012, alejandro_qml}, with
potential applications for exact and approximate
optimization, sampling, clustering, anomaly detection,
and simulating quantum many-body systems for
materials science and chemistry. Quantum computing's potential has also drawn interest
from other governmental institutions such as the
National Academy of Sciences \cite{NationalAcademiesofSciencesandMedicine2019} and the Executive
branch of the U.S. Government\cite{NationalStrategicOverview}. Quantum computing is currently in an era of noisy (error-prone), intermediate-scale ($<$100 qubits) quantum devices or \lq NISQ \rq \cite{Preskill_2018}, and developers have focused on the most promising near-term quantum computing application areas of chemistry / physical simulation, optimization, and machine learning.

\subsection{Quantum machine learning}

The field of quantum machine learning (QML) seeks to improve upon
machine learning algorithms using quantum
computing, which may be quantum variants of classical approaches or wholly novel algorithms. For example, researchers have leveraged quantum
computing approaches to enhance support vector machines for
handwriting recognition \cite{PhysRevLett.113.130503} and train fully-connected Boltzmann machines for sampling purposes \cite{arbitrary_pairwise, Henderson2019}.
Another recent QML algorithm is a quantum version of a CNN called the
quanvolutional neural network (QNN), which provides a new and potentially powerful modeling technique \cite{qnn}. While the QNN algorithm approach using "quantum convolutional" or "quanvolutional" layers is explained in great detail in \cite{qnn}, we will briefly summarize the approach and argument for a quantum advantage here.

The QNN approach transforms input images, represented mathematically as a $N$-by-$N$-by-$M$ tensor, where $N$ is the pixel width and height of the square image and M is the number of channels. A QNN can be considered simply as a CNN extended by adding one or more new layers composed of \textit{quanvolutional filters}. A quanvolutional filter acts just as a convolutional filter, processing sub-samples of these input tensors. Each pixel of the input sub-sample is mapped to a different variational parameter in a structured quantum circuit; we can name these pixel values $\theta$. We can represent a single quanvolutional filter $i$ operating on input $\theta$ as:
\begin{equation}
|\Psi_{i}\rangle = U_{i}(\theta) \left|0\right\rangle^{\otimes n}
\end{equation}
wherein $|\Psi_{i}\rangle$ is the output quantum state of quanvolutional filter $i$ operating on input $\theta$, n is the number of qubits in the circuit, and $U_{i}$ is the unitary matrix representing the quantum circuit transformation, which contains gates that have been parameterized by $\theta$ (rotational angle parameterization is an update to the original method; see \ref{qnn_improvements} for additional details). This output quantum state captures the extracted “feature information”, which is decoded using a user-defined decoding protocol function $d$ to generate the final output feature $o_{i}$:
\begin{equation}
o_{i} = d(|\Psi_{i}\rangle)
\end{equation}
wherein each $o_{i}$ value is a scalar output, just as a single scalar output is produced for each convolutional filter $j$ running onto the same input sub-samples $\theta$.

As laid out in \cite{qnn}, the argument for a potential quantum advantage is based on the following assumptions:

\begin{enumerate}
    \item The quantum transformations executed on classical input data by the quantum device are computationally intractable to calculate (or simulate) classically.
    \item The quantum transformations provide a benefit over other similar, tractable, classical transformations.
\end{enumerate}

This quanvolutional neural network (QNN) approach was first tested on the MNIST data set \cite{mnist}. Experiments showed that while the quantum features generated did provide a benefit over a classical convolutional neural network (CNN) without the same quantum transformations, it did not outperform a CNN with an additional layer of \mbox{classically-tractable} \mbox{non-linear} transformations. We conducted further experiments to determine potential quantum advantage \cite{Brett2019}, implementing a classification task on satellite images of the DeepSat data set \cite{deepsat}. Using a layer of 3-by-3 quanvolutional filters (requiring nine qubits with the basis state encoding protocol specified in \cite{qnn} on a single color channel), the QNN trained faster but converged on a lower test set accuracy than the comparative classical CNN.

\section{Materials and Methods}

\subsection{Caveats and objectives}
Before examining our experimental details, it is worth briefly noting the objectives of both the QNN approach and the DeepSat modeling problem in this work. First, while the eventual goal of QML research is to show a clear quantum advantage over purely classical techniques, the state of technology is still nascent. Considerable effort and research is still required to incrementally better understand algorithmic performance and optimize the various components in the entire stack. This paper endeavours to improve on the shortcomings of the original QNN approach \cite{qnn} and show improvements in terms of accuracy and processing capability (larger tensors), with the goal of illuminating a path forward to potential quantum advantage. Second, in terms of the machine learning application we selected to study, we deliberately chose a relatively standard image classification problem. While this does not include the data processing particulars that go into a complete space application pipeline, as well as the more specific end-user applications such as accelerating disaster responses \cite{disaster}, or identifying maritime objects from space \cite{dlr119201}, CNNs are fundamental components of these pipelines. For simplicity and more direct comparison to QNNs, we therefore chose a more straightforward image classification task with a standard CNN implementation, recognizing that if there is a benefit for this type of task, a similar benefit should be realized in a more complex stack.

\subsection{Data set}

In this experiment we benchmark the QNN using low-resolution
satellite imagery to demonstrate the utility of such
technology in this field. Specifically, all experiments were trained using the freely available SAT-4 data set \cite{Basu2015}. SAT-4 is a collection of 500,000 28-by-28 pixel 4-channel patches labelled as 4 distinct
classes – barren, trees, grassland and other.

In order to reduce the size of the available data and demonstrate the utility of both machine learning and quantum machine learning techniques in
aerospace, we use only 10,000 of the available patches (9,000 for training, and 1,000 for testing).

\subsection{Neural network architecture}
We compared a CNN and QNN model, both of which are comprised of an assortment of convolutional (CONV), pooling (POOL), and fully-connected (FC) layers. The QNN model also contains one quanvolutional (QUANV) layer. The structure of the CNN model was CONV1-POOL1-CONV2-POOL2-FC. Both CONV layers use a stride of 1, while the POOL layers have a stride of 2. CONV1 uses five $5\times5$ filters, while CONV2 uses twelve $3\times3$ filters. POOL1 is an average pooling layer with a $5\times5$ filter, and POOL2 is a maxpooling layer with a $2\times2$ filter. The QNN model follows a similar architecture. The only modification is the replacement of the layer CONV1 and POOL1 with a quanvolutional layer QUANV1. 

\subsection{Potential QNN improvements} \label{qnn_improvements}

Our analysis of our earlier experiments and their overall performance identified potential avenues for improvement for the QNN model approach in \cite{qnn}.

\begin{enumerate}
    \item \underline{Preserving more classical information.} Our previous method of basis state encoding required that each pixel value is thresholded to 0 or 1; this loses considerable information from the  input data. An encoding protocol which preserves more of this information is desirable.
    \item \underline{Processing larger input tensors.} While 3-by-3 pixel filters were sufficient for the MNIST data set, this input size was too small to properly capture input features in the DeepSat data set.  This limitation was due to both the encoding and decoding protocols of \cite{qnn}. The encoding basis state protocol required the same number of qubits as pixels, while the decoding protocol required the full probability distribution of all possible basis states. This limited us to 3-by-3 filters as simulating 25 qubit systems (which would be required for a 5-by-5 pixel input) would have taken a prohibitively long time. An alternate approach is needed to implement larger quanvolutional filters efficiently.
    \item \underline{Using structured rather than random circuits.} In our original work \cite{qnn}, we showed that adding more quanvolutional filters did increase accuracy for the QNN networks as expected (see Figure 2). However, adding many more quanvolutional filters did not increase accuracy results dramatically; this is likely due to the  construction of the quanvolutional circuits implementing their transformations. The circuits were simply different random quantum circuits, formed by randomly applying 1 and 2 qubit gates to the qubits in the circuit. As the random quantum circuit becomes sufficiently deep, the resulting output distribution approaches that of the Haar measure \cite{rqd}. This means in practice, although the circuits all had very different random gates choices, they likely were extracting very similar features from the input data. Selecting the quantum circuits more methodically should increase the variety of information the quanvolutional filters extract.
    \item \underline{Large pre-processing requirements.} Running large numbers of quantum circuit simulations is computationally expensive and slow. Rather than do these calculations on the fly, in \cite{qnn} a large amount of pre-processing was undertaken to calculate all possible output values for each possible basis state encoding for each quanvolutional filter circuit. While this was computationally feasible for 9 qubit circuits, it is grossly inefficient at scale. An alternate approach to pre-processing may improve the scalability.
\end{enumerate}

\subsection{Improved QNN approach} \label{improved_qnn_approach}
Our revised approach here aims to address each of these avenues for potential improvement with a new quanvolutional layer processing structure. The quanvolutional framework still has the same components as shown in Figure 1B of \cite{qnn}, namely (1) encoding, (2) quantum circuit processing, and (3) decoding protocols. We implemented changes in each of these protocols as follows:

\begin{itemize}
    \item \underline{Encoding protocol: variational parameters.} Instead of using basis encoding and discretizing all input pixels, we choose to preserve the pixel information by encoding them as continuous values into rotational angles using \mbox{1-qubit} gate rotations.
    \item \underline{Quantum circuit ansatz: non-random, structured circuits.} Figure \ref{fig:aspen25} shows the quantum computer used in this work, Rigetti’s Aspen-7-25Q-B quantum processing unit (QPU), which has 25 qubits with 24 programmable two-qubit gate interactions. This fit nicely with the number of parameters in our 5-5-4 tensor blocks of input data (see Figure \ref{fig:tensor-example}), which when flattened have 100 parameters. By generating graphs with the same topology as the Aspen-7-25Q-B and with random edge weights, we were able to create natively executable quantum approximate optimization algorithm (QAOA) \cite{Farhi2014c} circuits. These circuits assume that our randomly-weighted graphs with Aspen topology were MaxCut instances to solve, and returned solutions. We assigned a variational parameter for each term in the cost Hamiltonian, which maps to each edge in the graph, thus leading to 24 parameters. Additionally, after applying the cost Hamiltonian terms, we have an additional variational parameter for each application of the driver Hamiltonian, leading to 25 total parameters per $p$ layer in the QAOA design (see Appendix \ref{Appendix} for additional details on QAOA circuit interactions). If $p = 4$, there is exactly 1 variational parameter in the circuit for each input tensor pixel. This allows a natural way of mapping larger tensors into structured quantum circuits. However, we only ran the $p = 1$ case to attempt to circumvent gate depth issues (see Section \ref{new_qnn_challenges} for more details).

\begin{figure}
    \centering
    \includegraphics[width=\textwidth]{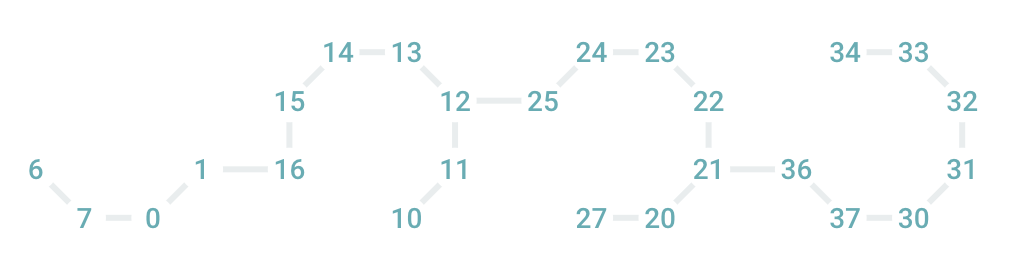}
    \caption{Visualization of the Rigetti Aspen-7-25Q-B QPU topology. Each number represents a programmable qubit, and edges represent the ability to run two-qubit gates between those qubits. By counting, one can verify this device has 25 qubits and 24 two-qubit gates.}
    \label{fig:aspen25}
\end{figure}

    \item \underline{Running on real QPU hardware.} Simulating a single shot of a 25+ qubit system can take several minutes, while on a real QPU device this process is on the order of microseconds. While still limited in terms of available access time due to other users reserving the QPU, we were able to run orders of magnitudes more results than would have been possible within the same block of time using simulations.

    \item \underline{Decoding protocol.} Instead of pre-computing all possible input/output basis state pairs, which is intractable at scale, we developed a dynamic mapping protocol. It works as follows: within a particular compute window, we run as many input tensors as possible through the QPU and generate output for each of these inputs. After the window is closed, we take all the input tensors which were successfully processed and use them to construct a balltree data structure \cite{balltree}. This then acts as a fast “map”, so that for a new, unseen tensor (i.e. one that was unprocessed by the QPU during the time available) we can quickly “map” this tensor to the most similar tensor that was successfully processed. The QPU output of this tensor is then used as an approximation for the unprocessed tensor. This method allows for a dynamic mapping regardless of the exact number of input tensors that were processed for each quanvolutional filter circuit.
\end{itemize}

\begin{figure}
    \centering
    \includegraphics[width=\textwidth]{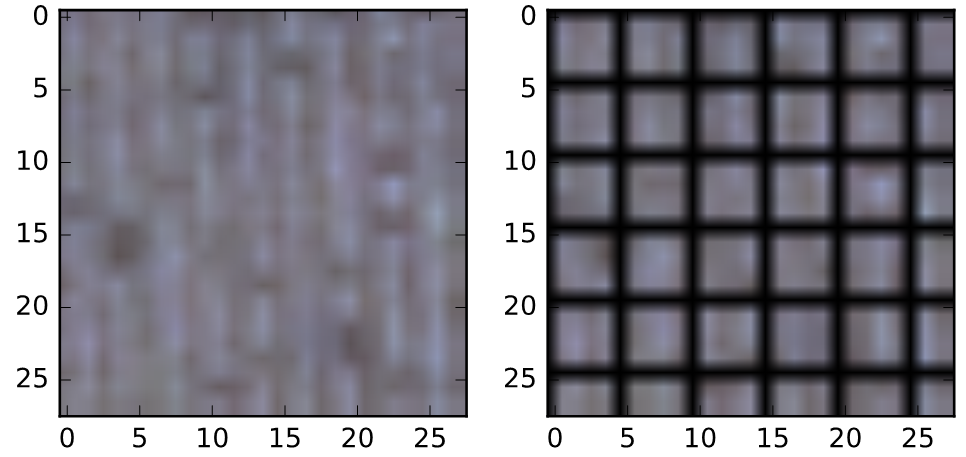}
    \caption{Example DeepSat-4 image (left) broken into 5-5-4 tensor blocks (right). Each of the complete 5-5-4 tensors blocks (25 per image) are sent as tensors to be run through the QPU circuit.}
    \label{fig:tensor-example}
\end{figure}

\section{Results}

\begin{figure}
    \centering
    \includegraphics[width=0.9\textwidth]{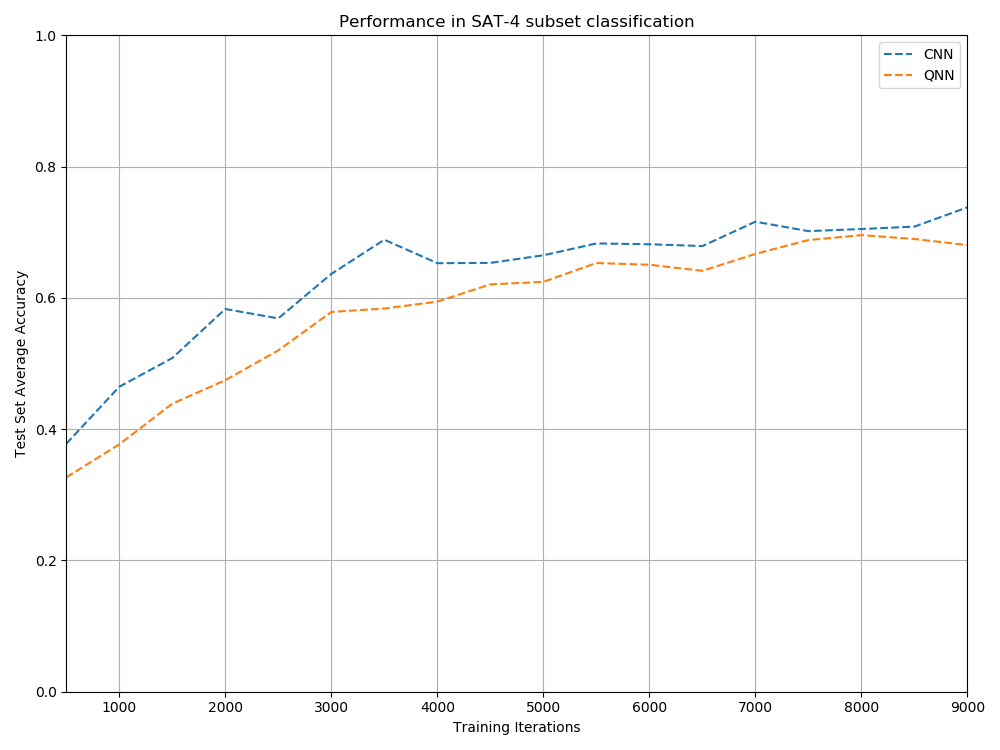}
    \caption{Results comparing two neural networks: the first being a QNN with an initial quanvolutional layer composed of 5 quanvolutional QAOA circuits and the second being a CNN with an additional CONV + POOL with 5 convolutional filters. The plotted results are the average across 10 networks, for each of CNN and QNN models.}
    \label{fig:results}
\end{figure}

Our experimental results are shown in Figure \ref{fig:results}, which take the average test set accuracy results as a function of training iterations across 10 different CNN and QNN models, respectively. As expected, for both the CNN and QNN models, test set accuracy increased as a function of training iterations. While the average CNN model results were more accurate than the QNN results, they converged to similar values towards the end of training with both models getting to around 70\% accuracy. While not yet outdoing the classical model, these results indicate that the new QNN protocol described in \ref{improved_qnn_approach} has significantly increased performance. In the previous results exploring the performance of the original QNN algorithm \cite{qnn} on the same DeepSat data modeling problem \cite{Brett2019}, the QNN model test set accuracy did not reach 60\% accuracy using 25 quanvolutional filters. These results show the improved QNN reaching approximately 70\% accuracy with only 5 quanvolutional filters.

\section{Discussion} \label{discussion}
Near-term quantum computing is a constant challenge. Dealing with one limitation almost inevitably unearths some new challenge. While the improvements to the QNN algorithm in this work have shown benefits over the original implementation of \cite{qnn}, there are still inefficiencies that could be further improved. To improve the algorithm, we believe that we need to address the challenges raised by the new method.

\subsection{New QNN challenges and potential solutions} \label{new_qnn_challenges}

\begin{itemize}
    \item \underline{Determine ideal number of QPU shots per input.} While simulators give exact quantum states, real QPU hardware provides information one measurement (‘shot’) at a time, providing at best an approximate quantum state. More shots give a better representation of the true quantum state after applying the quantum circuit in question.  In our experiments it is unclear how performance is affected by the number of shots taken in the quantum circuit. This could be explored experimentally in the future to optimize performance vs run-time.
    
    \item \underline{Improving heuristic dynamic mapping.} While computationally tractable, our decoding protocol essentially does a nearest-neighbor approximation of what the quantum output would be if that input tensor was run through a particular quanvolutional filter. The assumption that this approximation is sufficiently close to the value that would be returned by running the actual input through the QPU needs to be tested, and its impact on performance quantified. In particular, the heuristic dynamic mapping takes longer (1) as the input tensors grow larger and (2) the more quantum runs which are completed (i.e. more possible outputs to map to). To make this overall algorithm as strong as possible, the fastest, most scalable classical mapping to find the nearest processed tensor needs to be investigated. This could also involve some form of GPU parallelization as the nature of the underlying problem (calculating distances between many vectors) is highly parallelizable.
    
    \item \underline{Mitigating gate depth limitations.} While the speed benefit of running on real QPU hardware is impressive compared to simulation times, noise becomes a major factor when working with real quantum circuits. Error in quantum circuits both (1) corrupts results and consequently (2) limits total gate depth (i.e. number of gate operations while still being under some total error $\epsilon$). Error increases exponentially as a function of circuit depth for real quantum circuits in the absence of error-correction, so even near-perfect hardware will become error-dominant when running a sufficiently deep circuit. If we used the original $p = 4$ QAOA mapping, our total circuit depth would be roughly 150, which would result in extremely noisy output. We reduced the gate depth by grouping our pixels into groups of 4 and taking the average across those 4 for each encoding parameter in the $p = 1$ QAOA case, which led to a gate depth of around 40. However, this reduces the potential expressible power of the full $p = 4$ circuit considerably. To get better results and use deeper circuits, there are essentially two lines of improvement. First are changes made at the hardware level: each improvement in noise performance in the QPU hardware should have a positive impact on QNN performance. Second are solutions at the software level: by using various forms of quantum error correction protocols \cite{Erhard2019}, we may be able to improve the overall performance and explore deeper circuit ansatz.

    \item \underline{Experimenting with different structured quantum circuits.} In this study, we saw good modeling using variations of the same structured QAOA circuit ansatz. However, there are other types of ansatz that could be analyzed for particular data sets. This could lead to better understandings for optimal design choice of the number and type of quanvolutional filter circuits.

\end{itemize}

\section{Conclusion}
Quantum computing proffers a powerful alternative to purely classical computing methods for machine learning applications. While not yet showing a practical advantage by outperforming a classical state-of-the-art nonlinear layer (CONV + POOL), our new QNN algorithm demonstrates progress by building off our earlier work \cite{qnn}. This work provides more insight into how the quanvolutional approach might be applied more effectively in terms of accuracy and run-time efficiency. Such an iterative process, wherein each improvement in the overall QNN algorithm should raise the bar and help clarify the next tangible goals to improve, suggests that the road to quantum advantage will be incremental, not sudden. Our follow on work will continue this and try to work in the potential improvements mentioned in Section \ref{discussion} as we continue to reach towards a quantum machine learning implementation that shows a clear quantum advantage.

\section{Acknowledgements}
We would like to thank Duncan Fletcher for his time spent reviewing this work and helpful edits.

\bibliography{refs} 
\bibliographystyle{unsrt}

\begin{appendices}
\section{QAOA algorithm details} \label{Appendix}
All quanvolutional filters in this experimental study were QAOA ansatz, each solving a different randomly weighted graph. 

The QAOA algorithm is a universal gate-based quantum computing analogue to running the quantum annealing algorithm. Quantum annealing is used to solve quadratic, unconstrained binary optimization (QUBO) problems, of which many important NP-complete problems (such as MaxCut) can be expressed. In this section, we deeply explore the most trivial two-qubit case to understand the effects of the algorithm.

Quantum annealing seeks to evolve an initial ground state $\left|\Psi_{0}\right\rangle$ of an initial trivial Hamiltonian, to the ground state $|\Psi\rangle$ of a problem or cost Hamiltonian $H_P$. This can be expressed via the equation:

\begin{equation}
|\Psi\rangle= e^{-i H_{P} t}\left|\Psi_{0}\right\rangle
\end{equation}

where t is time. The QAOA algorithm as a quantum circuit emulates this process by essentially breaking it into 3 steps, which we will focus on for 2 qubits:

\begin{equation}
H_{I}=H \otimes H
\end{equation}

\begin{equation}
H_{P'}= Cnot \cdot\left(I \otimes R_{z}(\theta)\right) \cdot Cnot \approx e^{-i H_{P}}
\end{equation} 

\begin{equation}
H_{D'}=\left(I \otimes H R_{Z}(\beta) H\right) \cdot\left(H R_{Z}(\beta) H \otimes I\right) \approx e^{-i \beta H_{D}}
\end{equation}

where $H_{I}$ is the \textit{Initialization Hamiltonian}, $H_{P'}$ is the approximate \textit{Problem} or \textit{Cost Hamiltonian} term, and $H_{D'}$ is the approximate \textit{Driver} or \textit{Mixer Hamiltonian} evolved for a duration proportional to $\beta$, where:
\begin{center}
$H=\frac{\sqrt{2}}{2}\left[\begin{array}{cc}
{1} & {1} \\
{1} & {-1}
\end{array}\right]$,
$\quad \text{ $Cnot$ }=\left[\begin{array}{cccc}
{1} & {0} & {0} & {0} \\
{0} & {1} & {0} & {0} \\
{0} & {0} & {0} & {1} \\
{0} & {0} & {1} & {0}
\end{array}\right]$,
\end{center}

\begin{center}
$ \quad I=\left[\begin{array}{cc}
{1} & {0} \\
{0} & {1}
\end{array}\right]$,
$\quad R_{z}(\theta)=\left[\begin{array}{cc}
{\frac{-\theta}{e^{2}}} & {0} \\
{0} & {e^{\frac{\theta}{2}}}
\end{array}\right]$.
\end{center}
The $H_{P'}$ transformation is simply a decomposition of the ZZ-Ising gate:

\begin{equation}
    Z Z(\theta)=\left[\begin{array}{cccc}
{e^{\frac{\theta}{2}}} & {0} & {0} & {0} \\
{0} & {e^{\frac{-\theta}{2}}} & {0} & {0} \\
{0} & {0} & {e^{\frac{-\theta}{2}}} & {0} \\
{0} & {0} & {0} & {e^{\frac{\theta}{2}}}
\end{array}\right]
\end{equation}

except for a phase shift:

\begin{equation}
Cnot \cdot\left(I \otimes R_{z}(\theta)\right) \cdot Cnot =\left[\begin{array}{cccc}{e^{\frac{-\theta}{2}}} & {0} & {0} & {0} \\ {0} & {e^{\frac{\theta}{2}}} & {0} & {0} \\ {0} & {0} & {e^{\frac{\theta}{2}}} & {0} \\ {0} & {0} & {0} & {e^{\frac{-\theta}{2}}}\end{array}\right]
\end{equation}

\begin{figure}
    \centering
    \includegraphics[width=0.8\textwidth]{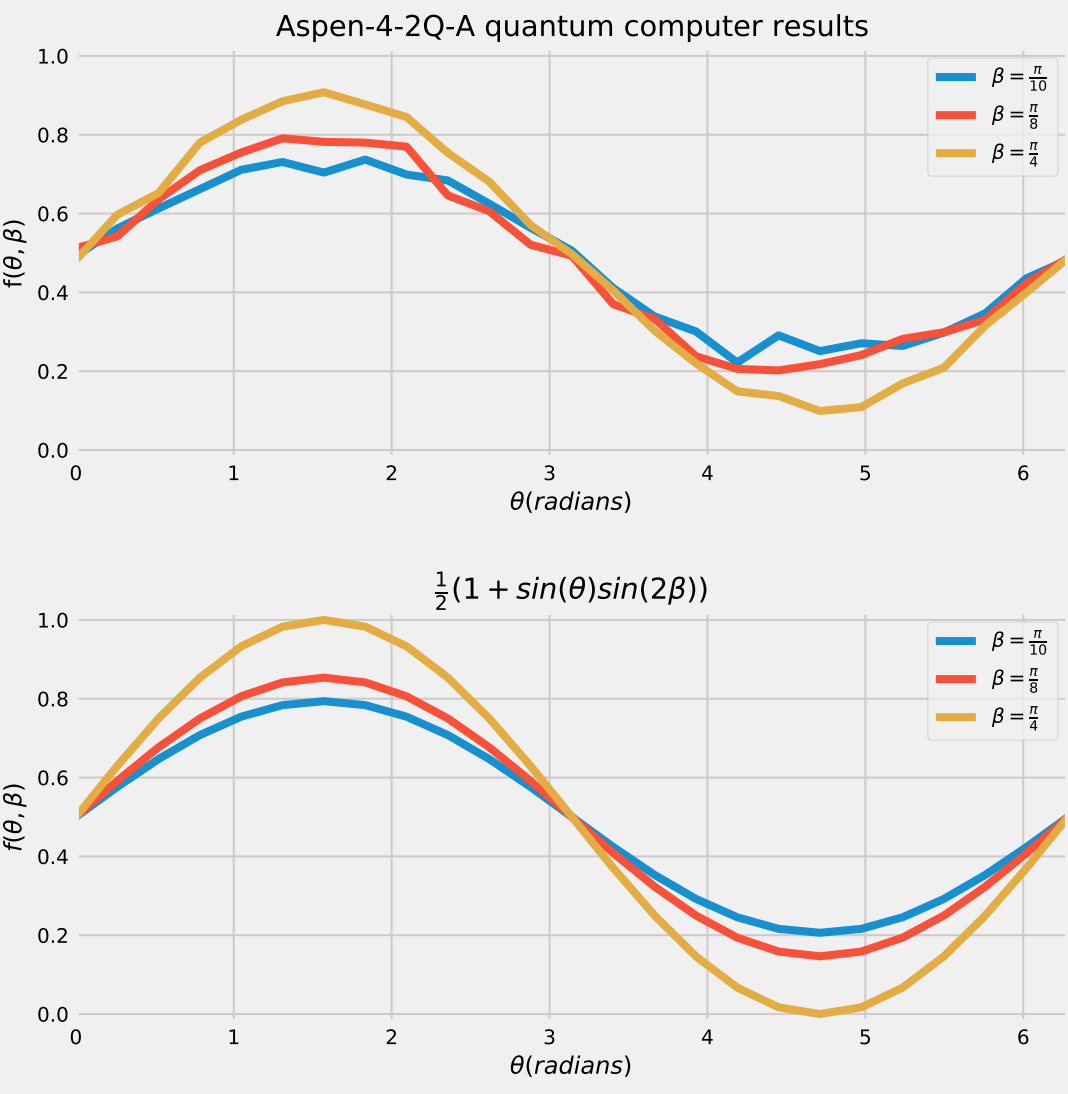}
    \caption{Visualizing the effect of a ZZ gate controlling the coupling strength between two qubits ran on actual quantum hardware (top) vs the exact analytic solution (bottom).}
    \label{fig:zz_results}
\end{figure}

By running our full circuit on an initial ground state of $|00\rangle,$ we get:
\begin{equation}
     H_{D'} H_{P'} H_{I}|00\rangle=|\Psi\rangle.
\end{equation}
Using $|\Psi\rangle$, we can calculate the exact probabilities of measuring if the qubits are in the same state (i.e. both in $|00\rangle$ or $|11\rangle$) as: 
\begin{equation}
    f(\theta, \beta)=\left\langle\Psi_{00} | \Psi_{00}\right\rangle+\left\langle\Psi_{11} | \Psi_{11}\right\rangle=\frac{1}{2}(1+\sin (\theta) \sin (2 \beta))
\end{equation}
with clear maximum when $\left.\sin (\theta) \sin (2 \beta)=1 \text { (i.e. } \theta=\frac{\pi}{2}+2 \pi n \text { and } \beta=\frac{\pi}{4}+n \pi\right)$ and minima
when $\sin (\theta) \sin (2 \beta)=-1$ (i.e. either $\theta=\frac{3 \pi}{2}+2 \pi n$ and $\beta=\frac{\pi}{4}+n \pi$ or $\theta=\frac{\pi}{2}+2 \pi n$ and $\beta=$
$\left.\frac{3 \pi}{4}+n \pi\right)$. This can be visualized in Figure \ref{fig:zz_results} for three different $\beta$ angles and a range between 0 and $2\pi$ for $\theta$.

\end{appendices}

\end{document}